\documentstyle[epsf,12pt]{article}

\begin{document}

\begin{flushright}
CEBAF-TH-95-16 \\
November 1995\\
hep-ph/9511272
\end{flushright}
\vspace{2cm}
\begin{center}
{\Large \bf Quark-Hadron Duality and Intrinsic
Transverse Momentum}\footnotemark
\end{center}
\begin{center}{A.V. RADYUSHKIN}\footnotemark
\\
{\em Physics Department, Old Dominion University,}
\\{\em Norfolk, VA 23529, USA}
 \\ {\em and} \\
{\em Continuous Electron Beam Accelerator Facility,} \\
 {\em Newport News,VA 23606, USA}
\end{center}
\footnotetext{Presented at XXXV  Cracow School
of Theoretical Physics, Zakopane  (Poland), June 4 - 14, 1995}
\footnotetext{Also at the Laboratory of Theoretical Physics,
JINR, Dubna, Russian Federation}

\vspace{2cm}

\begin{abstract}

It is demonstrated that local quark-hadron duality prescription
applied to several exclusive processes  involving the pion,
is equivalent to using an  effective
$ \bar q q $ (two-body) light-cone wave function
$\Psi^{(LD)}(x,k_{\perp})$ for the pion. This wave function
models    soft dynamics
of all higher $\bar qG \ldots G q $
Fock components of the standard light-cone approach.
Contributions corresponding to
higher Fock components in a hard regime
appear in this approach as radiative corrections
and are suppressed by powers of $\alpha_s/\pi$.

\end{abstract}
\vspace{1cm}

\newpage

\section{Introduction}

Study of effects due to  finite sizes of the
hadrons and incorporation of the
transverse momentum degrees of freedom
is a notoriously difficult  problem for the QCD analysis of
inclusive  (see, $e.g.,$ \cite{ht}) and exclusive  processes.
Since the advent of the parton model \cite{feynman},
it is taken for granted that the
hadron can be viewed as  a collection of quarks and gluons,
each of which carries a finite fraction $x_i P$ of the
large ``longitudinal'' momentum $P$ of the hadron,
and also some ``transverse'' momentum $k_{i \perp}$.
However,  a  justification of such a picture from the basic
principles of QCD is not a straightforward exercise.
To begin with,
the coordinate-representation version
of $k_{\perp}$ is a derivative $\partial_{\perp}$
in the transverse  direction.
In a gauge theory, $\partial_{\perp}$
always comes together with the gauge field $A_{\perp}$ in the
form of a covariant derivative
$  \partial_{\perp} \to D_{\perp}  = \partial_{\perp} -ig A_{\perp}$,
$i.e.,$ the finite-size effects are mixed with those
due to extra gluons.

In the  operator product expansion (OPE) approach
\cite{ope},
the nonperturbative aspects of the
hadron dynamics are described/parameterized
by matrix elements of local operators.
In particular, the
longitudinal momentum distribution is
related to the lowest-twist composite
operators  in which all the covariant
derivatives appear in traceless-symmetric
combinations
$ \{ D_{\mu_1} D_{\mu_2}
\ldots D_{\mu_n} \}$ \cite{gp,lnc,ar77}.
To take into account  transverse-momentum effects,
one  may wish to   consider matrix elements of higher-twist composite
operators in which some of the  covariant derivatives
appear in a
contracted  form like
$D^2 = D_{\mu} D^{\mu}$.
Attempting  to relate them to
transverse-momentum distributions
 (cf. \cite{eric}), one would immediately notice, however, that
$D^2 $ looks more like  analogue
of the quark virtuality $k^2$. Furthermore, the
presence of the gluonic
field $A_{\mu}$ in the covariant derivative
obscures such an  interpretation as well.
In particular,  using the equation of motion
$\gamma^{\mu} D_{\mu} q = 0$,
one can convert a two-body quark-antiquark operator
$  \bar q \{ \gamma_{\mu_1} D_{\mu_2} \ldots D_{\mu_n} \}
D^2 q$
with extra $D^2$
into the ``three-body''  operator
$  \bar q \{ \gamma_{\mu_1} D_{\mu_2} \ldots D_{\mu_n} \}
(\sigma^{\mu \nu} G_{\mu\nu}) q$ with  an  extra  gluonic field
 $G_{\mu\nu}$.
Moreover,  the two contracted
covariant derivatives $D_{\mu} \ldots D^{\mu}$
can be  separated by  $D_{\mu_k}$'s  forming  the traceless
combination and, to put $D_{\mu}$ and   $D^{\mu}$
next to each other, one should
perform commutations, each producing a $G$-field again \cite{dergepol}.
By choosing an optimized basis,
one can avoid some of the complications \cite{ht,jaffe},
but  the observations listed  above clearly show
that the OPE-inspired approach
is unlikely to produce
 a simple and intuitively
appealing basis for constructing
phenomenological
functions describing   the transverse-momentum
degrees of freedom.

Still, the OPE approach  has many  evident bonuses.
In particular, it is based on a covariant perturbation
theory in 4 dimensions and provides an
explicitly gauge-invariant
and   Lorentz-covariant description.
In this respect, it is  analogous
to  the Bethe-Salpeter formalism for  bound states.
However, the  well-known problem of
the Bethe-Salpeter formalism is the presence of the
unphysical variable of the relative time.
This variable is unnecessary
and, without a loss of information,
 one can describe bound states
in a 3-dimensional formalism.
This is achieved
by projecting the Bethe-Salpeter amplitude on
a particular
($e.g.,$ equal-time or light-front) hyperplane.

In the light-cone approach (see, $e.g.,$ \cite{bl80}),
a hadron  is described by a set of light-cone wave
functions (Fock components) $\Psi^{(N)}( \{ x_i, k_{\perp_i}  \})$.
For mesons, the two-body  wave function
$\Psi^{(2)}(x, k_{\perp})$ can be related to  the
Bethe-Salpeter amplitude
taken in the light-cone gauge
and integrated over the minus component
of the relative momentum \cite{blpi79}.
A bonus of the light-cone approach is that
effects due to  transverse momenta
are unambigously  separated
from those related to higher Fock components.
Hence, if the lowest Fock component
gives the   dominant contribution,
one can hope to construct
a reasonable phenomenology
based on  modelling  the two-body LC wave function.
On the other hand,  if  one should  perform a
summation over all Fock components,
the  predictive power of the scheme
is rather limited,
since  constructing models for numerous higher Fock components
leaves too much freedom for  the model builders.
This poses  a serious problem
for phenomenological applications
of the standard light-cone formalism.
In particular, if one uses the  popular BHL
set of constraints  for
the pion LC wave function
\cite{bhl},
the $\bar q q$ component  always
contributes less than 50\%
to the pion form factor value at $Q^2=0$,
and  higher Fock components are
absolutely needed to ensure the  correct normalization
$F_{\pi}(Q^2=0) =1$.

A possible way out is to introduce an effective two-body
wave function (see, $e.g., $\cite{schlumpf}),
which  includes
the low-energy contribution
from the higher Fock components,
so that one would get
$F_{\pi}(Q^2=0) \approx 1$
 just from the overlap
of these  wave functions.
One can interpret such a wave function
as a wave function for a constituent quark,
$i.e.,$ a quark dressed by soft  gluons.
However,  for  different  processes,
the higher Fock components can
appear with different process-dependent
weights, and it is not clear {\it a priori}
whether  the effective wave function
can  be introduced in a universal way.
Another point is   that while
absorbing information about  soft gluons
into a $\bar q q$ wave function
may be a good approximation, hard
gluons cannot  be correctly
taken into account in this way.
Hence, even using the effective two-body wave function,
one should allow for a possibility of
having explicit multi-body wave functions.
  Still, the dominant role
of the effective two-body component may take place
in such a  scheme, since  each emission of a  hard gluon  is
suppressed by the QCD coupling constant
$\alpha_s/\pi \sim  0.1$, and   the contribution
of the multi-body  components
may  be relatively small.
The  problem is that it is unclear
how to combine the QCD corrections
with  the constituent quark picture,
because the constituent quark is
not a field one readily finds in
the original QCD Lagrangian.

Here, using the pion as an
example,  we will outline a new approach
to transverse-momentum effects
in exclusive processes. It is  based on
QCD sum rule ideas.
On several examples, we show that  results obtained
using the quark-hadron duality
prescription   \cite{nr82} can be reformulated in terms
of a universal effective wave function
$\Psi^{LD}(x,k_{\perp})$ absorbing information
about  soft dynamics.
The scheme  starts with diagrams of
ordinary covariant perturbation
theory and  allows for a systematic inclusion
of the radiative corrections
in a way totally consistent with the basics of QCD.

\section{Handbag diagram  and $\xi$-scaling}

\setcounter{equation} 0

A naive idea is that, to take into account
effects due to the finite size of the hadrons,
one should just write the ``parton model''
formulas without neglecting intrinsic  transverse momentum
in hard scattering amplitudes.
In doing  this, however, one should  explicitly specify
a field-theoretic  approach which is used for such a
generalization of the standard
parton model.
As emphasized in the Introduction,
one can choose  here between at least two basically different
alternatives: standard covariant 4-dimensional formalism
or 3-dimensional approaches analogous to the old-fashioned
perturbation theory.
The bonus of the 4-dimensional
  approach, in the form of the  OPE, is  a gauge-invariant
and a Lorentz-covariant  description of the hadrons
in terms of matrix elements of composite operators.
However,  as argued above, interpretation of the OPE results in terms of
 transverse degrees of freedom is not self-evident.
Moreover, there are some
practically important
amplitudes which are ``protected''  from the
dynamical $(D^2)^n$-type
higher-twist corrections.
The most well-known example  is given by
the classic ``handbag'' diagram for deep inelastic
scattering.
As we will see below, in a scalar toy model
its contribution
contains only target-mass corrections,
$i.e.$, it gives no information about finite-size effects.
In QCD, the handbag contribution
 contains  a twist-4 operator with extra $D^2$,
but no operators with higher powers of $D^2$.

\subsection{Scalar model}

\begin{figure}[t]
\mbox{
   \epsfxsize=12cm
 \epsfysize=4cm
  \epsffile{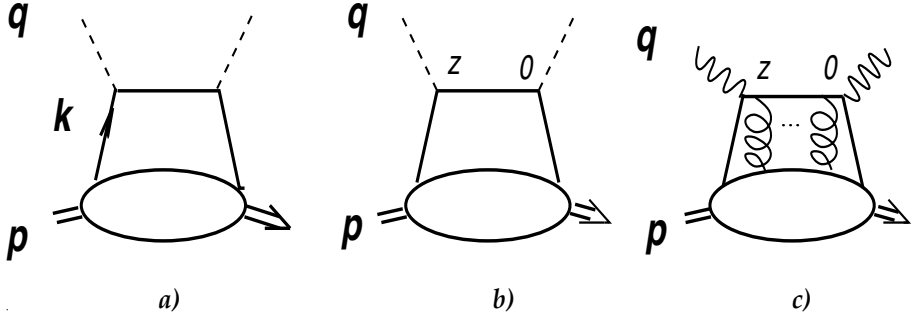}  }
  \vspace{1cm}
{\caption{\label{fig:1} Handbag diagram.
{\it a)} Momentum representation for  scalar model.
{\it b)} Coordinate representation for  scalar model.
{\it c)} QCD modification of the quark propagator.
   }}
\end{figure}

To  illustrate the effect  in its cleanest form,
let us consider the handbag contribution in  a model
where all fields are scalar (Fig.\ref{fig:1}):
\begin{equation}
T(q,p) = - \int \frac{d^4 k}{(k+q)^2} F(p,k).
\end{equation}
At large $Q^2 \equiv -q^2$, one can neglect the parton virtuality
$k^2$ in $(k+q)^2= q^2 +2(kq)+k^2$
and expand the  propagator in powers of $2(kq)/q^2$  to obtain
\begin{equation}
T(q,p) = \int d^4 k \sum_{n=0}^{\infty}
 \frac{ (2qk)^n}{Q^{2n}}  F(p,k).
\end{equation}
Now, the integral
\begin{equation}
\int  k^{\mu_1} \ldots k^{\mu_n} F(p,k) d^4 k =
A_n  p^{\mu_1} \ldots p^{\mu_n}  + traces \
\end{equation}
is evidently the matrix element of a local operator
with $n$ derivatives.The usual parton density $f(x)$ is introduced
 by treating the coefficients $A_n$ as its moments
\cite{gp,lnc}:
\begin{equation}
A_n = \int_0^1 x^n f(x) dx.
\end{equation}
As a result, the amplitude can be written as
\begin{eqnarray}
T(q,p) = \frac1{Q^2} \int_0^1 \sum_n^{\infty}
\left [ \frac{2x (qp)}{Q^2} \right ]^n f(x) dx + O(1/Q^4) \nonumber \\ =
\int_0^1 f(x) \frac1{ Q^2 - 2x(qp) } + O(1/Q^4) \ .
\end{eqnarray}
Taking its imaginary part $W(q,p) \sim \, $Im$ \, T(q,p)$,  we get
\begin{equation}
W(q,p) =
\int_0^1 f(x) \, \delta ( Q^2  - 2x(qp) ) \, dx + O(1/Q^4)
= \frac1{Q^2} x_{B} f (x_{B}) + O(1/Q^4) ,
\label{eq:Bj}
\end{equation}
where $x_{B} \equiv Q^2/2(qp)$ is the standard Bjorken variable.

\subsection{Power corrections and $\xi$-scaling}

Eq.(\ref{eq:Bj})  gives the lowest-twist contribution.
The power-suppressed terms denoted by $O(1/Q^4)$
are apparently due to the neglected $k^2$ term in the original propagator.
One can expect that supplementing the $(kq)/Q^2$ expansion by
the  $k^2/Q^2$ expansion, one can take into account the effects due to
nonzero virtuality $k^2$ by introducing phenomenological
functions related to matrix elements like
 \begin{equation}
\int  k^{\mu_1} \ldots k^{\mu_n} (k^2)^N F(p,k) d^4 k \,=
A^{(N)}_{n}  p^{\mu_1} \ldots p^{\mu_n} + traces \ .
\end{equation}
Of course, one should be more careful now with  the ``traces''
in this parameterization.
The best way to
maintain the necessary accuracy is well-known: one should take
the traceless part  $\{k^{\mu_1} \ldots k^{\mu_n}\}$
of the original tensor $k^{\mu_1} \ldots k^{\mu_n}$.
Then,  the right-hand-side will also be a traceless tensor,
constructed from the 4-vector $ p^{\mu}$.

So, if one decides to keep the $k^2$ terms,
one  should supplement this
by a re-expansion of the  $(kq)^n$-factors
over  traceless tensors.
 In fact, the actual problem is simpler than it seems, because
a straightforward  expansion
of the propagator is just  in terms of the traceless
combinations:
\begin{equation}
- - \frac{1}{(q+k)^2} \biggr |_{k<q} =
\sum_{n=0}^{\infty}  \frac{2^n}{(Q^2)^{n+1}}
q^{\mu_1} \ldots q^{\mu_n} \{k_{\mu_1} \ldots k_{\mu_n} \}.
\label{eq:ex}
\end{equation}
Note, that  there are no
$\{k_{\mu_1} \ldots k_{\mu_n} \} (k^2)^N$ terms with $N \neq 0$
in this expansion: the $(k^2)^N$ terms from a naive
expansion over powers of $(kq)$ and $k^2$
are exactly cancelled by $(k^2)^N$ terms from the reexpansion
of $(qk)^n$-factors over traceless tensors.
In other words,  the handbag diagram is
insensitive to  nonzero-virtuality effects.
Introducing the twist-2 distribution function $via$
\begin{equation}
 \int   \{k^{\mu_1} \ldots k^{\mu_n} \} F(p,k) d^4k
= \{p^{\mu_1} \ldots p^{\mu_n} \} \int_0^1 x^n f (x) dx
\end{equation}
and  performing the summation over $n$ by inverting the expansion formula
(\ref{eq:ex}),
we get
\begin{equation}
T(q,p) =
\int_0^1 \frac1{ (q + xp)^2 } f(x) \, dx \, .
\end{equation}
The essential point is that no power-suppressed terms were neglected
in this derivation. Hence, we can write
$(q + xp)^2 = -Q^2 + 2x(qp) +x^2 p^2$ keeping all the terms
here  and
calculate the imaginary part:
\begin{equation}
W(q,p) =
\int_0^1 f(x) \, \delta \left (- Q^2  + 2x(qp) +x^2 p^2 \right ) dx
= \frac1{Q^2} \frac{x_{B}} {{\sqrt{1+ \frac{4p^2 x_{B}^2}{Q^2} }}}
\,  f (\xi) ,
\end{equation}
where
\begin{equation}
\xi = \frac{2 x_{B}} {1+ \sqrt{1+\frac{4p^2 x_{B}^2}{Q^2}} }
\end{equation}
is the Nachtmann-Georgi-Politzer $\xi$-variable \cite{nacht},
\cite{gp}.
Hence, all the power-suppressed contributions contained
in the handbag diagram can be interpreted as the target-mass
corrections. In particular, the handbag contribution
 contains   no power
corrections for a massless target.

\subsection{Coordinate representaion}

Absence of the higher twist terms as well as the
possibility to easily calculate  the target-mass dependence
of the handbag contribution is directly related to the fact
that the  propagator of a massless particle
has a simple singularity structure.
To illustrate this, let us write $T(q,p)$
in the coordinate representation:
\begin{equation}
T(q,p) = \int \frac{d^4z}{z^2} \, e^{iqz} \,
\langle  p | \phi (0)  \phi(z) | p \rangle .
\label{eq:Fscalar}
\end{equation}
The first term in the $z^2$-expansion for  the matrix element
\begin{equation}
 \langle  p | \psi (0)  \psi (z) | p \rangle
=  \xi_2(zp) + z^2 \xi_4(zp) + (z^2)^2 \xi_6 (zp) + \ldots
\label{eq:z2expansion}
\end{equation}
corresponds to the twist-2 distribution amplitude
\begin{equation}
\chi (zp)\biggl |_{z^2=0} = \int_0^1 \varphi (x) e^{ix(zp)} \, dx
\biggr |_{z^2=0} \, ,
\label{eq:twist2wf}
\end{equation}
while subsequent  terms correspond to operators containing an
increasing number of $\partial^2$'s.
It is straightforward to observe that,
while the twist-2 term produces the $1/Q^2$ contribution,
the twist-4 term
is accompanied by an  extra $z^2$-factor which
completely kills the $1/z^2$-singularity of the quark propagator,
and the result of the $d^4z$
integration is proportional in this case to $\delta^4(q-xp)$,
$i.e.,$  this term
is invisible for large $Q^2$. The same is evidently true for all the terms
accompanied by higher powers of $z^2$.
This means that the handbag diagram contains only one
term:  it cannot generate
higher powers of $1/Q^2$  which one could interpret as
the $(\langle k^2 \rangle/Q^2)^n$  expansion.

For spin-1/2 particles, the quark propagator
$S^c(z) \sim \gamma_{\mu} z^{\mu}/(z^2)^2$
has a stronger singularity for $z^2 =0$,
which is cancelled only by the $O(z^4)$
term in the expansion of the matrix element
$\langle  p | \bar q(0) \gamma_{\mu}
 q(z) | p\rangle $.
Hence, one may expect that  there is
 a non-vanishing  twist-4
contribution corresponding to  the $O(z^2)$  term
of this expansion,  but no higher terms.
In a gauge theory, like QCD, one should also take into account
the fact that the gluonic  field $A_{\nu}$,
in a covariant gauge, has zero twist.
As a result, if the  gluons have longitudinal
polarization, the   configurations  shown in Fig.\ref{fig:1}$c$   are
not power-suppressed compared to the original
handbag contribution.
The net result of
such gluonic insertions into
the  quark propagator is a phase factor
\begin{equation}
 S^c(z) \to S^c(z) P e^{igz^{\nu} \int_0^1 A_{\nu}(tz) dt} \{ 1 + O(G) \} ,
\end{equation}
where the $O(G)$ term corresponds to
insertion of physical gluons. The latter are described
by the gluonic field-strength tensor $G_{\alpha \beta}$
and  produce $1/Q^2$-suppressed contributions.
Thus,  including the phase factor, we get the modified
QCD handbag contribution
\begin{equation}
T (q,p) \sim  \int d^4z \, e^{-iq_1z} \,
\frac{z^{\mu}}{(z^2)^2}
\langle  p | \bar q(0) \gamma_{\mu}
e^{igz^{\nu} \int_0^1 A_{\nu}(tz) dt} q(z) | p \rangle .
\label{eq:fQCDcoord}
\end{equation}
The matrix element
\begin{equation}
\langle  p | \bar q(0) \gamma_{\mu}
e^{igz^{\nu} \int_0^1 A_{\nu}(tz) dt}
q(z) | p \rangle
\label{eq:}
\end{equation}
can be Taylor-expanded just like
$\langle  p | \bar q(0) \gamma_{\mu} q(z) | p \rangle$,
with the only change $\partial_{\nu} \to D_{\nu}$
in the resulting local operators.
Thus, the  incorporation of gauge invariance
does not change our conclusion
that the (generalized) handbag diagram  cannot generate  a tower of the
$(1/Q^2)^n$  corrections  which one could interpret as
the $(\langle k_{\perp}^2 \rangle/Q^2)^n$
or $(\langle k^2 \rangle/Q^2)^n$ expansion.
The power corrections are produced by the final-state interaction
which is described by complicated contributions due to the
operators of
$ \bar q G \ldots G q $-type.
At twist 4, the $\bar q  D^2 q$ and  $ \bar q G q $
terms combined together can be interpreted in terms
of functions related to operators of $\bar q  D_{\perp}^2 q$
type \cite{ht}. However, for twist-6 and higher, the absence of
the generic $\bar q  (D^2)^N q$ contribution stops
further progress in this direction.
Hence, a  simple phenomenological  description
of higher-twist  corrections in terms of something like the
transverse-momentum distribution $f(x,k_T)$ is impossible.

\section{Exclusive processes:
 $\gamma^* \gamma^* \to \pi^0$  transition}

 \setcounter{equation} 0

The transition  $\gamma^*(q_1) \gamma^*(q_2) \to \pi^0(p)$
of two virtual photons into
a neutral pion (Fig.\ref{fig:2}$a$) is the cleanest
exclusive process  for testing QCD predictions.
 The relevant form factor
$F_{\gamma^* \gamma^* \pi^0} \left(q^2,Q^2\right)$,
with  $q^2 \equiv -q_1^2, Q^2 \equiv -q_2^2$, can be defined
  in terms of the pion-to-vacuum matrix element
of the product of two electromagnetic currents:
\begin{equation}
4\pi \int d^4x\,e^{-iq_1x}
\langle {\pi},\stackrel{\rightarrow}{p}
|T\left\{J_{\mu}(x)\,J_{\nu}(0)\right\}| 0 \rangle
 = i \sqrt{2}  {\epsilon}_{{\mu}{\nu}{q_1}{q_2}}\,
F_{\gamma^* \gamma^* \pi^0}\left(q^2,Q^2\right).
\label{eq:form}
\end{equation}

\subsection{pQCD results}

\begin{figure}[t]
\mbox{
   \epsfxsize=12cm
 \epsfysize=4cm
  \epsffile{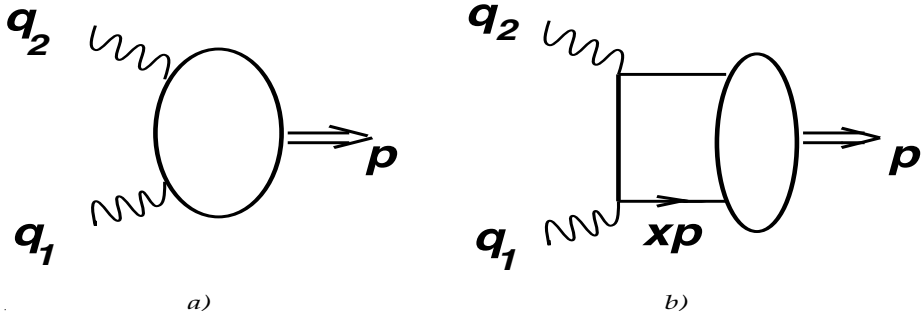}  }
  \vspace{1cm}
{\caption{\label{fig:2}
Form factor of the $\gamma^* \gamma^* \to \pi^0$ transition. {\it a)}
General structure.
{\it b)} Leading-order pQCD term.
   }}
\end{figure}

In the lowest order of
perturbative QCD, the asymptotic behaviour of
$F_{\gamma^* \gamma^* \pi^0} \left(q^2,Q^2\right)$
can be calculated  from a diagram
similar to the handbag diagram (see Fig.\ref{fig:2}$b$) \cite{bl80}.
The basic change is that one should
use now the pion distribution amplitude
$\varphi_{\pi}(x)$ instead of the parton density $f(x)$:
\begin{equation}
F_{\gamma^* \gamma^*  \pi^0 }^{pQCD}(q^2,Q^2) = \frac{4\pi}{3}
\int_0^1 {{\varphi_{\pi}(x)}\over{xQ^2+\bar x q^2}} dx  +
O(\alpha_s/\pi) + O(1/Q^4).
\label{eq:g*g*pipqcd}
\end{equation}
Experimentally,
the most important situation is when one of the photons
is real: $q^2=0$. In this case, pQCD predicts that \cite{bl80}
\begin{equation}
F_{\gamma \gamma^*  \pi^0 }^{pQCD}(Q^2) = \frac{4\pi}{3}
\int_0^1 {{\varphi_{\pi}(x)}\over{xQ^2}} dx  +O(\alpha_s/\pi) + O(1/Q^4).
\label{eq:gg*pipqcd}
\end{equation}
The  nonperturbative information is accumulated here by
the integral
\begin{equation}
I = \int_0^1 {{\varphi_{\pi}(x)}\over{x}} dx \
\label{eq:I }  .
\end{equation}
Its value depends on the shape of the
pion distribution amplitude $\varphi_{\pi}(x)$.
In particular,  using  the
asymptotic form \cite{tmf78,pl80,blpi79}
\begin{equation}
\varphi_{\pi}^{as}(x) = 6 f_{\pi} x(1-x)
\label{eq:phias}
\end{equation}
gives the following prediction for the large-$Q^2$
behaviour \cite{bl80}:
\begin{equation}
F_{\gamma \gamma^*  \pi^0 }^{as}(Q^2) = \frac{4 \pi f_{\pi}}{Q^2}.
\label{eq:fggpias}
\end{equation}

\subsection{Anomaly and BL-interpolation}

Of course, the asymptotic $1/Q^2$-dependence cannot
be the true behaviour
of $F_{\gamma \gamma^*  \pi^0 }(Q^2)$  in the low-$Q^2$
region,  since  the $Q^2=0$ limit of
$F_{\gamma \gamma^*  \pi^0 }(Q^2)$
is known to be finite and
 normalized by the $\pi^0 \to \gamma \gamma$ decay rate.
In fact,  incorporating PCAC and ABJ anomaly \cite{anomaly},
one can calculate $F_{\gamma \gamma^*  \pi^0 }(0)$
theoretically:
\begin{equation}
F_{\gamma \gamma^*  \pi^0 }(0) = \frac1{\pi f_{\pi}}.
\label{eq:anomaly}
\end{equation}
In pQCD, one can imagine that the transition
from the high-$Q^2$ asymptotics to the
low-$Q^2$ behaviour is reflected
by higher twist corrections of $(M^2/Q^2)^n$-type,
which may sum up into something  like $1/(Q^2+M^2)$,
$i.e.,$ some   expression finite at $Q^2=0$ and behaving like $1/Q^2$
for large $Q^2$.
This idea was originally formulated
 by Brodsky and Lepage \cite{blin}
who proposed the   interpolation formula
\begin{equation}
F_{\gamma \gamma^*  \pi^0 }(Q^2) =
{ {1} \over {\pi f_{\pi}
\left (1+{{Q^2}\over{4 \pi^2 f_{\pi}^2}} \right )}},
\label{eq:blin}
\end{equation}
which reproduces both the $Q^2 =0 $ value (\ref{eq:anomaly})
and the high-$Q^2$ behaviour  (\ref{eq:fggpias}) with
the normalization
 corresponding to
the asymptotic distribution amplitude (\ref{eq:phias}).

The BL-interpolation formula (\ref{eq:blin}) has a  monopole form
$$F_{\gamma \gamma^*  \pi^0 }(Q^2) \sim 1/(1+Q^2/s_0)$$ with
the scale $s_0 = 4 \pi^2 f_{\pi}^2 \approx 0.67 \, GeV^2$,
which is  numerically   close to the $\rho$-meson mass squared:
$m_{\rho}^2 \approx 0.6 \, GeV^2$.
So, the BL-interpolation suggests a form similar to that based on
the VMD expectation
$F_{\gamma \gamma^*  \pi^0 }(Q^2) \sim 1/(1+Q^2/m_{\rho}^2)$.
In the VMD-approach,   the $\rho$-meson mass $m_{\rho}$
serves as a parameter  determining  the pion charge radius,
and one can
expect that the tower of $(s_0/Q^2)^N$-corrections
suggested by the BL-interpolation formula can be attributed to
the intrinsic transverse momentum.

\subsection{Light-cone formalism and power corrections}

As  noted earlier,  the relative weight and
 interpretation of  power  corrections
depends on a particular formalism
used for a beyond-the-leading-twist extension
of pQCD formulas.
In the operator product expansion  approach,
the lowest-order (in $\alpha_s$) ``handbag'' contribution to the
$\gamma \gamma^* \to \pi^0$ form factor  again has only
pion-mass corrections:
\begin{equation}
F_{\gamma \gamma^* \pi^0}^{handbag} \left(Q^2\right) = \frac{4\pi}{3}
\int_0^1 {{\varphi_{\pi}(x)}\over{xQ^2+x(1-x)m_{\pi}^2}} \, dx .
\end{equation}
This means that, in the OPE approach, the $(``s_0"/Q^2)^N$-type
corrections  can come only from
the $\bar q G \ldots G q$ operators,
for which  simple phenomenology is impossible.

As an alternative to the covariant
perturbation theory and OPE,
one can use the light-cone (LC) formalism \cite{bl80}, in
which effects  due to the
intrinsic transverse momentum $k_{\perp}$
are  described by
the  light-cone wave function  $\Psi(x,k_{\perp})$.
The LC   formula for the $\gamma \gamma^* \to \pi^0$
form factor  looks like
\begin{equation}
(\epsilon_{\perp} \times q_{\perp}) F_{\gamma \gamma^* \pi^0} (Q^2) \sim \int
\Psi(x,k_{\perp})
\frac{(\epsilon_{\perp} \times (xq_{\perp}-k_{\perp})) }
{ (xq_{\perp}-k_{\perp})^2} dx d^2 k_{\perp} \, ,
\label{eq:FscalLC}
\end{equation}
where $q_{\perp}$ is a two-dimensional vector in the transverse plane
satisfying $q_{\perp}^2=Q^2$ and $\epsilon_{\perp}$ is a vector orthogonal to
 $q_{\perp}$ and also lying in
the transverse plane \cite{bl80}.
In the LC  formalism,  quark and gluon fields
are on-shell. However, the invariant
mass ${\cal M}$ of an intermediate state
does not coincide with the pion mass.
In particular,  for the $\bar qq$-component,
${\cal M}^2 = ( k_{\perp}^2+m_q^2)/x(1-x)$. Hence,
integrating over  $k_{\perp}$ is equivalent to integrating
over the invariant masses (or LC-energies) of intermediate states, with
$\Psi(x,k_{\perp})$ specifying the  probability amplitude
for each particular ``mass''.

Unfortunately,
eq.(\ref{eq:FscalLC})  also has little chances
of producing the series of the
$\langle k_{\perp}^2 \rangle/Q^2$ corrections, because
the  expansion
\begin{equation}
\frac{1}{(xq_{\perp}-k_{\perp})^2} = \frac{1}{x} \sum_0^{\infty}
2^n \left [ \frac{\theta(|k_{\perp}|<xQ)}{(xQ^2)^{n+1}}
+ \frac{\theta(|k_{\perp}|>xQ)}{(k_{\perp}^2/x)^{n+1}} \right ]
q_{\perp}^{\mu_1} \ldots q_{\perp}^{\mu_n} \{k_{\perp}^{\mu_1}
\ldots k_{\perp}^{\mu_n} \}
\label{eq:legendre}
\end{equation}
contains only traceless combinations $\{k_{\perp}^{\mu_1}
\ldots k_{\perp}^{\mu_n} \}$.
Substituting it into eq.(\ref{eq:FscalLC}) and using that
the wave function  $\Psi(x,k_{\perp})$ depends   on $k_{\perp}$
only through $k_{\perp}^2$, we obtain that
all terms of this expansion (proportional to Legendre's polynomials)
vanish after the angular integration, except
for the $n=0$ and $n=1$ terms.
As a result,   the leading $1/Q^2$ term is corrected
only by the term  resulting
from integration of $\Psi(x,k_{\perp})/k_{\perp}^2$ over the region
$|k_{\perp}|> xQ$:  there is no  tower of
$(\langle k_{\perp}^2 \rangle/Q^2)^N$
power corrections  in the handbag contribution.
Thus, one is forced again to explain the $(1/Q^2)^N$-corrections
by contributions from  higher $\bar q G \ldots G q$ Fock components,
which unavoidably leads  to a complicated phenomenology.

The last but not the least comment is that   the LC formalism
is based on  bound-state equations like $\Psi = K \otimes \Psi$,
with $K$ being an ``interaction kernel''.
However, it is not clear whether such an equation
has any justification  in QCD outside perturbation theory.
Moreover,  it is known that QCD has a lot of nonperturbative
effects: complicated vacuum, quark and gluon condensates, $etc.,$
which play a dominant role in determining the
properties of QCD  bound states.
So, it is very desirable to develop   a QCD description
of  hadrons in terms of functions similar to  the
bound state wave functions $\Psi(x,k_{\perp})$,
but without assuming existence of   bound state equations.
Below, we outline our   attempt to derive such a description from
 QCD sum rules and quark-hadron duality.

\section{Basics of quark-hadron duality }

\setcounter{equation} 0

\subsection{Outline of the QCD sum rule calculation of $f_{\pi}$}

The basic idea of the QCD sum rule approach \cite{svz}
is the quark-hadron  duality,
$i.e.,$ the possibility to describe one and the same
object  in terms of either quarks or hadrons.
To calculate
$f_{\pi}$, we consider  the $p_{\mu}p_{\nu}$-part  of the
correlator of two axial currents:
\begin{equation}
\Pi^{\mu\nu}(p) =
i \int e^{ipx}  \langle 0 | T \,(j_{5\mu}^+(x)\,j_{5\nu}^-(0)
\,)|\,0\rangle\, d^4 x =
p_{\mu}p_{\nu}\Pi_2(p^2)-g_{\mu\nu}\Pi_1(p^2).
\label{eq:Pi}
\end{equation}
The dispersion relation
\begin{equation}
 \Pi_2(p^2)= \frac1{\pi}\int_0^{\infty}\frac{ \rho(s)}{s-p^2}ds +
``subtractions"
\label{eq:PiDR}
\end{equation}
represents $\Pi_2(p^2)$ as an integral  over hadronic spectrum
with the spectral
density $\rho^{hadron}(s)$  determined by projections
\begin{equation}
\langle 0 | j_{5\mu} (0) |\pi;  P \rangle = i f_{\pi} P_{\mu},
\label{eq:fpi}
\end{equation}
$etc.,$ of the axial current onto
hadronic states
\begin{equation}
\rho^{hadron}(s) = \pi f_{\pi}^2 \delta(s-m_{\pi}^2) + \pi f_{A_1}^2
\delta(s-m_{A_1}^2)  + ``higher \ \  states"
\label{eq:rhohadron}
\end{equation}
($f_{\pi}^{\exp} \approx 130.7 \, MeV$ in our normalization).
On the other hand, when the probing virtuality  $p^2$ is negative and large,
one can use the operator product expansion
\begin{equation}
\Pi_2(p^2) = \Pi_2^{pert}(p^2) + \frac{A}{p^4} \langle \alpha_s GG \rangle
+ \frac{B}{p^6} \alpha_s \langle \bar qq \rangle^2  + \ldots
\label{eq:PiOPE}
\end{equation}
where $\Pi_2^{pert}(p^2) \equiv \Pi_2^{quark} (p^2)$ is the
perturbative  version  of   $\Pi_2 (p^2)$ given by a sum of
pQCD Feynman diagrams while the condensate terms
$\langle GG \rangle$,  $\langle \bar qq \rangle$, $etc.,$
(with perturbatively calculable coefficients
$A, B, etc.$ )
describe/parameterize  the  nontrivial structure of the QCD vacuum.

For the quark amplitude $\Pi_2^{quark}(p^2)$,  one can also write down
the  dispersion relation  (\ref{eq:PiDR}),  with $\rho(s)$
substituted by its perturbative analogue $\rho^{quark}(s)$:
\begin{equation}
\rho^{quark}(s)= \frac1{4 \pi} \left ( 1 + \frac{\alpha_s}{\pi}
+ \ldots \right )
\label{eq:rhoquarkPi}
\end{equation}
(we neglect quark masses). Hence, for large $-p^2$, one can write
\begin{equation}
\frac1{\pi} \int_0^{\infty} {{\rho^{hadron}(s) - \rho^{quark}(s)}\over
{s-p^2}} ds \ = \frac{A}{p^4} \langle \alpha_s GG \rangle
+ \frac{B}{p^6} \alpha_s\langle \bar qq \rangle^2  + \ldots \ .
\label{eq:sumrule}
\end{equation}
This expression essentially states that the condensate
terms describe  the difference between  the
quark and hadron spectra.
At this point, using the known values of the condensates,
one can try to construct  a model for the hadronic spectrum.

In the axial-current channel, one has
an infinitely   narrow pion peak
$\rho_{\pi} = \pi f_{\pi}^2 \delta(s-m_{\pi}^2)$,
a rather wide  peak at
$s \approx 1.7 \, GeV^2$ corresponding to  $A_1$ and then
 ``continuum'' at higher energies. The simplest model
is to treat  $A_1$ also as a part  of the continuum,
$i.e.,$ to use  the model
\begin{equation}
\rho^{hadron}(s) \approx \pi f_{\pi}^2 \delta(s-m_{\pi}^2) + \rho^{quark}(s) \,
 \theta(s \geq s_0),
\label{eq:rhomodelPi}
\end{equation}
in which   all the higher resonances including the  $A_1$
are approximated  by the quark  spectral density  starting at some
effective threshold $s_0$.
Neglecting the pion mass and  requiring the best agreement between the two
sides
of the resulting  sum rule
\begin{equation}
{{f_{\pi}^2}\over{p^2}} = \frac1{\pi} \int_0^{s_0} {{\rho^{quark}(s)}\over
{s-p^2}} ds \ + \frac{A}{p^4} \alpha_s \langle GG \rangle
+ \frac{B}{p^6} \alpha_s \langle \bar qq \rangle^2 + \ldots \
\label{eq:fpisumrule}
\end{equation}
in the region of large $p^2$,
we can fit the remaining parameters  $f_{\pi}$ and  $s_0$
which specify  the model spectrum.
In practice, the more convenient SVZ-borelized version
of this sum rule (multiplied by $M^2$)
\begin{equation}
{f_{\pi}^2} = \frac1{\pi} \int_0^{s_0} \rho^{quark}(s) e^{-s/M^2} {ds}
 \  +\frac{\alpha_s\langle GG\rangle}{12\pi M^2}
		  +\frac{176}{81}\frac{\pi\alpha_s\langle\bar qq\rangle^2}{M^4}
		    + \ldots
\label{eq:fpisumruleborel}
\end{equation}
is used for actual fitting. Using the standard values for the condensates
$\langle GG \rangle$, $\langle \bar qq \rangle^2$,
we adjust $s_0$ to get an (almost) constant  result
for the rhs of eq.(\ref{eq:fpisumruleborel})
starting with the minimal possible value of the SVZ-Borel parameter $M^2$.
The magnitude  of   $f_{\pi}$ extracted in this way,  is very close
to its  experimental value $f_{\pi}^{exp} \approx 130 \, MeV.$

\subsection{Local duality}

Of course, changing the  values of the condensates,
one would get the best stability for a different magnitude
of  the effective threshold
$s_0$, and the resulting value of  $f_{\pi}$ would also change.
There exist  an evident  correlation
between the values of $f_{\pi}$ and $s_0$ since, in the
$M^2 \to \infty$ limit, the sum rule reduces to the local duality relation
\begin{equation}
f_{\pi}^2 = \frac1{\pi} \int_0^{s_0} \rho^{quark}(s) \, ds.
\label{eq:fpiLD}
\end{equation}

Thus, the local quark-hadron duality  relation  states that,
despite their absolutely different  form,
the two densities $\rho^{quark}(s)$ and $\rho^{hadron}(s)$
give the same result if one integrates
them over the $appropriate$ duality interval $s_0$.
The role of the condensates was to determine the size of the duality
interval $s_0$, but after it was fixed,
one can write down the relation (\ref{eq:fpiLD}) which does not
involve the condensates.

Using the explicit lowest-order expression
$\rho^{quark}_0(s) = 1/4\pi$, we get
\begin{equation}
s_0 = 4\pi^2 f_{\pi}^2.
\label{eq:s0}
\end{equation}
Notice that $s_0 = 4\pi^2 f_{\pi}^2$ is
exactly the combination which appeared
in the Brodsky-Lepage interpolation formula
(\ref{eq:blin}). Numerically, $4\pi^2 f_{\pi}^2 \approx 0.67 \, GeV^2$,
$i.e.,$ the pion duality interval  is  very close to the
$\rho$-meson mass: $m_{\rho}^2 \approx 0.6 \, GeV^2$.
In fact, in   the next-to-leading order
\begin{equation}
\rho^{quark}_{NLO}(s) = \frac1{4\pi}\left (1+\frac{\alpha_s}{\pi} \right ).
\end{equation}
So, using   $\alpha_s/\pi \approx 0.1$,
one gets $s_0$  practically coinciding with $m_{\rho}^2$.
For the form factors,  this leads  to results
close to  the VMD expectations, even though no explicit
reference to the existence of the $\rho$-meson  is made.

\subsection{Local duality and pion wave function}

\begin{figure}[t]
\mbox{
   \epsfxsize=8cm
 \epsfysize=4cm
  \epsffile{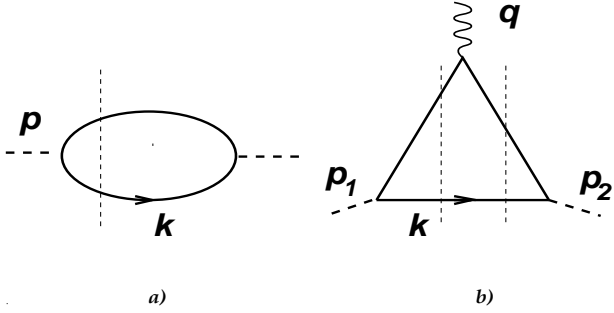}  }
  \vspace{1cm}
{\caption{\label{fig:3}
Leading-order  contributions for {\it a)}
two-point spectral density  $\rho(s)$ (4.14) and
{\it b)}
three-point spectral density
$\rho(s_1,s_2,Q^2)$ (5.8).  The narrow dashed lines indicate
Cutkosky cuts. }}
\end{figure}

In the  lowest order, the perturbative spectral density is given
by the Cutkosky-cut quark loop integral (see Fig.\ref{fig:3}$a$)
\begin{equation}
\rho^{quark}(s) = \frac{3}{2\pi^2} \int
\frac{k_+}{p_+} \left ( 1- \frac{k_+}{p_+}  \right )
\delta^{(+)} \left (k^2 \right )  \delta^{(+)} \left ( (p-k)^2 \right ) d^4k
\end{equation}
where $s\equiv p^2$.
Introducing the light-cone variables for $p$ and $k$:
$$p= \{p_+ \equiv P, p_- = s/P, p_{\perp} =0 \} \  ;  \
k = \{k_+\equiv xP, k_-, k_{\perp} \}$$
and integrating over $k_-$, we get
\begin{equation}
\rho^{quark}(s) = \frac{3}{2\pi^2} \int_0^1 dx \int
 \delta \left (s- \frac{k_{\perp}^2}{x \bar x} \right ) \, d^2 k_{\perp} .
\end{equation}
The delta-function  here   expresses the
fact that, since we are
working in the 4-dimensional formalism,
the light-cone combination
${{k_{\perp}^2}/{x \bar x}}$  coincides with $s$,
the square  of the external  momentum $p$.

Substituting now  $\rho^{quark}(s)$ into the local duality formula,
we obtain
\begin{equation}
f_{\pi}^2 =  \frac{3}{2\pi^3} \int_0^1 dx \int
\theta \left (k_{\perp}^2\leq  x\bar x s_0 \right ) \, d^2 k_{\perp} \, .
\label{eq:fpiLDkperp}
\end{equation}
This  representation
 has the structure similar to the expression for $f_{\pi}$
in the light-cone formalism \cite{bl80}
\begin{equation}
f_{\pi} = \sqrt{6} \,
\int_0^1 dx \int \Psi(x,k_{\perp}) \, \frac{d^2 k_{\perp}}
{8 \pi^3} \,
 ,
\label{eq:fpiLC}
\end{equation}
where $\Psi(x,k_{\perp})$ is the $\bar qq$-component of the
pion light-cone wave function.
To cast  the local duality result
(\ref{eq:fpiLDkperp}) into the form of eq.(\ref{eq:fpiLC}),
we introduce the ``local duality''
wave function for the pion:
 \begin{equation}
\Psi^{LD}(x,k_{\perp}) = \frac{2\sqrt{6}}{f_\pi} \,
\theta(k_{\perp}^2\leq  x\bar x s_0) \ .
\label{eq:PsiLD}
\end{equation}
 The specific form dictated by the local duality
implies that $\Psi^{LD}(x,k_{\perp})$
simply imposes a sharp cut-off at $ k_{\perp}^2x\bar x = s_0$.

It should be emphasized that in eq.(\ref{eq:fpiLC}) we
are integrating over $ k_{\perp}$,
$i.e.,$  the combination $ k_{\perp}^2x\bar x$ no longer
coincides with the mass$^2$ of the external particle
(which is $m_{\pi}^2$ in our case).
This is precisely the feature  of 3-dimensional bound-state
formalisms: internal particles are on mass shell $k^2=m_q^2$,
but off the energy shell:  $ k_{\perp}^2x\bar x \neq m_{\pi}^2$.
In what follows, we show, on less trivial examples,
that    the local duality
prescription allows one to get  results
reminiscent of 3-dimensional  formalisms,
though all the actual calculations are performed
in the standard 4-dimensional perturbation theory.

\section{Pion electromagnetic  form factor}

\setcounter{equation} 0

\subsection{Sum rule}

To demonstrate  that  the function   $\Psi^{LD}(x,k_{\perp})$
really has the properties one
expects from the pion  wave function, let us consider the quark-hadron
duality in a more complicated context  of
the pion electromagnetic  form factor $F_{\pi}(Q^2)$.
It is defined by
\begin{equation}
\langle p_2 | J^{\mu}(0) | p_1 \rangle = (p^{\mu}_1 + p_2^{\mu})F_{\pi}(Q^2),
\label{eq:piemf}
\end{equation}
where $Q^2 = -(p_2-p_1)^2$.
To apply the QCD sum rule technique,
we should consider in this case the correlator \cite{is82,nr82}
\begin{equation}
T^{\mu}_{\alpha\beta}(p_1,p_2) =
i \int e^{-ip_1x+ip_2y}
\langle 0| T \{j_{\beta}(y) J^{\mu}(0) j^+_{\alpha}(x)\}
|0 \rangle d^4x d^4y
\label{eq:piemcorr}
\end{equation}
of two axial currents $j_{\alpha}^+,j_{\beta} $ and one electromagnetic
current
$J^{\mu}$.
The pion EM form factor  can be extracted from the invariant amplitude
$T(p_1^2,p_2^2,Q^2)$
corresponding to the structure $P_{\alpha}P_{\beta}P^{\mu}$,
where $P=(p_1+p_2)/2$.

The obvious  complication now is that we have two channels to be
``hadronized",  since  the pion is present both in the initial
and final states. This necessitates the use of  the double dispersion relation
\begin{equation}
T(p_1^2,p_2^2,q^2)=\frac1{\pi^2}\int_0^\infty ds_1\int_0^\infty ds_2
 \ \frac{\rho(s_1,s_2,Q^2)}{(s_1-p_1^2)(s_2-p_2^2)}
+ ``subtractions"
\label{eq:doubledr}
\end{equation}
involving the double spectral  density
$\rho(s_1,s_2,Q^2)$.
Its hadronic version $\rho^{hadron}(s_1,s_2,Q^2)$
contains the term corresponding to the pion
form factor
\begin{equation}
\rho_{\pi\pi}(s_1,s_2,Q^2)= \pi^2f_{\pi}^2F_{\pi}(Q^2)
\delta(s_1-m_{\pi}^2)\delta(s_2-m_{\pi}^2)
\label{eq:rhopipi}
\end{equation}
and the contributions corresponding to transitions
between the pion and higher resonances, and also the terms related to
elastic and transition form factors of the higher resonances.
To construct the two-dimensional analog of the
``lowest state plus continuum'' ansatz,  we will  treat
all the contributions, except for the $\rho_{\pi\pi}$, as ``continuum'',
{\em i.e.}, we will model  $\rho^{hadron}(s_1,s_2,Q^2)$ by the
$\rho^{quark}(s_1,s_2,Q^2)$
outside the square $(0,s_0)\times (0,s_0)$:
\begin{equation}
\rho(s_1,s_2,Q^2) = \rho_{\pi\pi}(s_1,s_2,Q^2) +
\left(1- \theta(s_1<s_0)\theta(s_2<s_0)\right)\rho^{pert.}(s_1,s_2,Q^2).
\label{eq:rho2model}
\end{equation}
The SVZ-borelized sum rule (with $M_1^2=M_2^2\equiv M^2$)
for the pion form factor  then has the form  \cite{is82,nr82}
\begin{eqnarray}
f_\pi^2 F_\pi(Q^2)&=&\frac1{\pi^2}\int_0^{s_0} ds_1\int_0^{s_0} ds_2\
\rho^{quark}(s_1,s_2,Q^2)\exp\left(-\frac{s_1+s_2}{M^2}\right)
\nonumber \\
	     & & \hspace{20mm} +\frac{\alpha_s\langle GG\rangle}{12\pi M^2}
	         +\frac{16}{81}\frac{\pi\alpha_s\langle\bar qq\rangle^2}{M^4}
	             \left(13+\frac{2Q^2}{M^2}\right)
\label{eq:FpiSR}
\end{eqnarray}
(the pion mass was neglected as usual).

\subsection{Local duality}

In the large-$M^2$ limit, this gives  the local duality relation
\cite{nr82}
\begin{equation}
f_{\pi}^2F_{\pi}^{LD}(Q^2) = \frac1{\pi^2}\int_0^{s_0}ds_1
\int_0^{s_0} ds_2 \, \rho^{quark}(s_1,s_2,Q^2).
\label{eq:FpiLD}
\end{equation}

Again, the perturbative spectral density $\rho^{quark}(s_1,s_2,Q^2)$
corresponding to the triangle diagram Fig.\ref{fig:3}$b$
can be easily calculated
using the Cutkosky rules and light-cone variables in the frame where
the initial momentum $p_1$ has no transverse
components $p_1= \{p_1^+ = P, p_1^- = s_1/P,0_{\perp}\} $,
while the momentum transfer $q \equiv p_2-p_1$ has  no
``plus'' component: $p_2= \{P, (s_2+Q_{\perp}^2)/P, Q_{\perp}\} $:
\begin{equation}
\rho^{quark}(s_1,s_2,Q^2) = \frac3{2 \pi} \int_0^1 \,  dx \, \int d^2 k_{\perp}
\,  \delta \left (s_1 - {{k_{\perp}^2}\over{x \bar x}} \right )
\delta \left (s_2 - {{(k_{\perp}+xq)^2}\over{x \bar x}} \right ) \  .
\label{eq:rhoquarkFpi}
\end{equation}
Here, $x$ is the fraction of the total ``plus''
light-cone  momentum carried
by the  quark absorbing the momentum transfer
from  the virtual photon and  $k_{\perp}$
is its transverse momentum.

Substituting  $\rho^{quark}(s_1,s_2,Q^2)$
into the local duality relation, we get   the light-cone
formula for the pion form factor
\begin{equation}
F_{\pi}^{LD}(Q^2) = \int_0^1 \,  dx \, \int {{d^2 k_{\perp}}\over{16 \pi^3}} \,
\Psi^{LD} (x, k_{\perp}) \, \Psi^{LD} ( x, k_{\perp}+xq) ,
\label{eq:DYW}
\end{equation}
where $\Psi^{LD} (x, k_{\perp})$ is exactly the
local duality (\ref{eq:PsiLD}) wave function
introduced in the previous section.

At $Q^2=0$, the LC formula reduces to the   integral
\begin{equation}
P = \int \frac{dx\, d^2 k_{\perp}}{16\pi^3} |\Psi(x,k_{\perp})|^2
\label{eq:probab}
\end{equation}
which can be interpreted as  the probability
to find the pion in the state described by the
wave function $\Psi(x,k_{\perp})$.
Using the explicit form
of  $\Psi^{LD}(x,k_{\perp})$, we immediately get
$$P = \frac{s_0}{4 \pi^2 f_{\pi}^2},$$
which reduces to 1
if  we take the  lowest-order value $s_0 = 4 \pi^2 f_{\pi}^2$
 for the duality interval.
Hence,  $P^{(0)} =1$, $i.e.,$  the probability
to find the pion in the state described by
$\Psi^{LD}(x,k_{\perp})$ is 100\%.
In other words,
$\Psi^{LD}(x,k_{\perp})$  may  be treated
as an effective  wave function absorbing the
low-energy information about all Fock components.

It should be emphasized, however, that
if one uses the next-to-leading order value
$$s_0^{(1)} = {{4 \pi^2 f_{\pi}^2}\over{1+\frac{\alpha_s}{\pi}}}$$
for the duality interval, the probability integral $P$
will be smaller than 1 ($P^{(1)} \approx 0.9$ for  $\alpha_s \approx 0.3$).
This is a direct manifestation that the local duality
prescription explicitly produces contributions which can be interpreted
as  hard parts of the higher  Fock components like $\bar qGq, etc.$
(see Fig.\ref{fig:4} and Fig.\ref{fig:5} below).
However, the total probability to find the pion in such a higher Fock state is
rather small: it is suppressed by the factor
$\frac{\alpha_s}{\pi}\approx 0.1$ \, .

Though the probability integral  $P$
differs from 1 beyond the leading order,
the  relation  $F_{\pi}^{LD}(0) =1$
holds to all orders of perturbation theory.
The reason is that,  at any order,
there exists  the  Ward identity relation
between  the 3-point function $T^{\mu}_{\alpha \beta}(p,p)$ and
the 2-point function $\Pi_{\alpha \beta}(p)$:
$T^{\mu}_{\alpha \beta}(p,p) = - \partial \Pi_{\alpha \beta}(p)/ \partial
p_{\mu}$.
As  a result, the 3-point spectral density $\rho^{quark}(s_1,s_2,Q^2)$
reduces to the 2-point spectral density  $\rho^{quark} (s)$:
\begin{equation}
 \rho^{quark}(s_1,s_2,Q^2=0)  =
\pi \delta(s_1-s_2) \rho^{quark} (s_1).
\label{eq:WID}
\end{equation}
Hence, for $Q^2=0$, the duality integral  for the pion form factor
automatically reduces to a one-dimensional integral over $s_1$ (or $s_2$),
and the duality integral (\ref{eq:FpiLD}) for $f_{\pi}^2 F_{\pi}(0) $
would coincide with that
for  $f_{\pi}^2$ (\ref{eq:fpiLD}),  provided that the sides of the duality
square for the three-point function
are exactly equal to the duality
interval  $s_0$ for the two-point function  (the latter
not necessarily being equal to
the lowest-order value $s_0^{LO} = 4 \pi^2 f_{\pi}^2$).  As a result,
the local duality prescription  gives $F_{\pi}^{LD}(0)=1$
to all orders of perturbation theory.
In the lowest order, it also gives $P^{(0)}=1$.

One should not overestimate the accuracy of
the local duality results in the region of small $Q^2$.
Though   $F_{\pi}^{LD}(Q^2)$  dictates the values rather close in magnitude
to the VMD curve  $F_{\pi}^{VMD}(Q^2)$ or any other fit to data,
the LD-formula
\begin{eqnarray}
F_{\pi}^{ LD}(Q^2) =
1 - {{1+\frac{6s_0}{Q^2}}\over{\left ( 1+
 \frac{4s_0}{Q^2} \right ) ^{3/2} }  }
\label{eq:sqrt}
\end{eqnarray}
gives  infinite slope at $Q^2=0$,  and one should
not use it for calculating  the derivatives   of $F_{\pi}(Q^2)$
below $Q^2 \sim s_0$.
 As emphasized above, we obtained the correct value for $F_{\pi}(0)$
only because this value was protected by the Ward identity.
It is well known that the 3-point function in the small-$Q^2$
region has more complicated  quark-hadron
duality  properties which require a separate  study.
For the same reason, the local duality  fails
to produce reasonable valence parton densities for the pion.

\section{Quark-hadron duality for the
$F_{\gamma^* \gamma^* \pi^0} \left(Q^2\right)$
form factor}

\setcounter{equation} 0

\subsection{Basics}

    Within  the QCD sum rule
approach, one can  extract information about   the
$\gamma^* \gamma^* \to \pi^0$ form factor
from   the three-point correlation function  (\cite{NeRa83}):
\begin{equation}
{\cal{F}}_{\alpha\mu\nu}(q_1,q_2)= \frac{4\pi}{i \sqrt{2}}
\int d^4x\,d^4y\ e^{-iq_1 x- iq_2 y}
\langle 0 |T\left\{J_{\mu}(x)\,J_{\nu}(y)\,j_{5 \alpha}(0)\right\}| 0 \rangle
\label{eq:corr}
\end{equation}
calculated in the region where all the
virtualities \ $q_1^2 \equiv - q^2 ,q_2^2\equiv -Q^2$ and
$p^2\!=\!(q_1+q_2)^2$ are spacelike.

To study   the form factor $F_{\gamma^* \gamma^* \pi^0}(q^2,Q^2)$,
one should consider
the invariant  amplitude $F \left(p^2,q^2,Q^2\right)$ corresponding to the
tensor structure ${\epsilon}_{{\mu}{\nu}{q_1}{q_2}} p_{\alpha}$.
The dispersion relation
for the three-point  amplitude
\begin{equation}
F\left(p^2, q^2,Q^2 \right)={1\over{\pi}}\int_0^{\infty}
\frac{{\rho}\left(s,q^2,Q^2\right)}{s-p^2}\,ds
+ \mbox{ subtractions}
\label{eq:dispgg}
\end{equation}
specifies the relevant spectral density ${\rho}\left(s,q^2,Q^2\right)$.
Its hadronic version
\begin{equation}
{\rho}^{hadron}\left(s,q_1^2,q_2^2\right) = \pi f_{\pi}
F_{\gamma^* \gamma^* \pi^0}(q^2,Q^2)
\delta(s-m_{\pi}^2) + ``higher \ \  states "
\label{eq:rhohadrongg}
\end{equation}
contains the  term with the form factor we are interested in.
The relevant perturbative spectral density ${\rho}^{quark}(s,q^2,Q^2)$,
in the lowest order, is  given by  the integral representation
\begin{equation}
\rho^{quark}(s,q^2,Q^2)=2 \int_0^1
\delta \left ( s - \frac{q^2x_1x_3+Q^2x_2x_3}{x_1x_2} \right ) \
\delta(1-\sum_{i=1}^3 x_i)\
dx_1dx_2dx_3
\label{eq:rpts}
\end{equation}
in terms of the Feynman parameters for
the  one-loop triangle diagram.
 Scaling  the integration variables:
$x_1+x_2 =  y$, $x_2 =  xy$, $x_1 =(1-x)y \equiv \bar x y$
and taking trivial integrals over $x_3$ and $y$, we get
\begin{equation}
\rho^{quark}(s,q^2,Q^2)=2\int_0^1 \frac{x\bar{x}(xQ^2+ \bar x q^2)^2}
{[s{x}\bar{x}+xQ^2+ \bar x q^2]^3} \,dx  \, .
\label{eq:rhopt}
\end{equation}
It can be shown that the variable $x$
here is the light-cone fraction of the
pion momentum $p$ carried by one of the quarks.

\subsection{Local duality}

Incorporating the local duality, we obtain
\begin{equation}
 F_{\gamma^* \gamma^* \pi^0}^{LD}(q^2,Q^2)
= \frac1{\pi f_{\pi}} \int_0^{s_0} \rho^{quark}(s,q^2,\, Q^2)
=  \frac2{\pi f_{\pi}}
\int_0^1 dx \int_0^{s_0} ds \frac{x\bar{x}(xQ^2+ \bar x q^2)^2}
{[s{x}\bar{x}+xQ^2
+ \bar x q^2]^3}  \,  .
\label{eq:FLDggIntR}
\end{equation}
Substituting  the variable $s$
(the mass$^2$ of the $\bar qq$ pair)
 by the light-cone combination
$k_{\perp}^2/ x\bar x $, we get
$F_{\gamma^* \gamma^* \pi^0}^{LD}(q^2,Q^2)$
as an integral  over
the longitudinal momentum fraction
$x$ and the  transverse momentum $k_{\perp}$:
\begin{equation}
F_{\gamma^* \gamma^* \pi^0}^{LD}(q^2,Q^2) = \frac2{\pi^2 f_{\pi}}
\int_0^1 dx \, \int d^2k_{\perp} \, \theta(k_{\perp}^2\leq  x\bar x s_0)
\frac{(xQ^2+ \bar x q^2)^2}{ (xQ^2+k_{\perp}^2)^3}.
\label{eq:FLDxkper}
\end{equation}
Finally, introducing the effective wave function $\Psi^{LD}(x,k_{\perp})$
given by (\ref{eq:PsiLD})
we can  write $F^{LD}\left(Q^2\right)$  in the ``light-cone'' form:
\begin{equation}
 F_{\gamma^* \gamma^* \pi^0}^{LD}(q^2,Q^2) = \frac1{\pi^2 \sqrt{6}}
\int_0^1 dx \, \int d^2k_{\perp} \,\Psi^{LD}(x,k_{\perp})
\frac{(xQ^2+ \bar x q^2)^2}{ (xQ^2 + \bar x q^2+k_{\perp}^2)^3} \,  .
\label{eq:FLDLC}
\end{equation}
It is instructive to analyze this expression in some
particular limits.

\subsection{Limits}

{\it 1. Both photons are real.}
When both $q^2$ and  $Q^2$ are small, we can use the fact that
\begin{equation}
\frac{\mu^4}{ (\mu^2+k_{\perp}^2)^3} \to \frac1{2} \,  \delta (k_{\perp}^2)
\label{eq:delta2}
\end{equation}
in the $\mu^2 \to 0$ limit,  to obtain
(cf.\cite{bhl}) that the $\pi^0 \to \gamma \gamma$
decay rate is determined by the magnitude of the pion wave function
at zero transverse momentum:
\begin{equation}
F_{\gamma^* \gamma^* \pi^0}^{LD}(0,0) = \frac1{2 \pi \sqrt{6}} \int_0^1 \,
\Psi^{LD}(x,k_{\perp}=0)  \, dx .
\label{eq:FLD(0)}
\end{equation}
Using the explicit form of $\Psi^{LD}(x,k_{\perp})$
(\ref{eq:PsiLD}), we obtain
\begin{equation}
 F_{\gamma^* \gamma^* \pi^0}^{LD}(0,0) = \frac1{\pi f_{\pi}},
\end{equation}
which is exactly the value (\ref{eq:anomaly}) dictated by the axial anomaly.

{\it 2.  pQCD limit.} Assuming that  both $q^2$ and  $Q^2$ are so large
that the $k_{\perp}^2$-term can be neglected, we get the expression
\begin{equation}
F_{\gamma^* \gamma^* \pi^0}^{LD}(q^2,Q^2) = \frac1{\pi^2 \sqrt{6}}
\int_0^1 \frac{dx}{xQ^2+ \bar x q^2} \,
\int d^2k_{\perp} \,\Psi^{LD}(x,k_{\perp}) + O(1/Q^4) \,  .
\label{eq:FLDLChigh}
\end{equation}
Identifying the wave function
integrated over the transverse momentum
with  the pion distribution amplitude
\begin{equation}
\varphi_{\pi}^{LD}(x) = {{\sqrt{6}}\over{(2\pi)^3}}
\int \Psi^{LD}(x,k_{\perp}) \, d^2k_{\perp} \, ,
\end{equation}
we arrive at the lowest-order pQCD formula
\begin{equation}
F_{\gamma^* \gamma^*  \pi^0 }^{pQCD}(Q^2) = \frac{4\pi}{3}
\int_0^1 {{\varphi_{\pi}(x)}\over{xQ^2+\bar x q^2}} dx  +O(1/Q^4)
\label{eq:g*g*pipqcd2}
\end{equation}
for the large-virtuality  behaviour of the $\gamma^* \gamma^* \to \pi^0$
transition form factor.

{\it 3. One real photon.}
A very simple result for $\rho^{quark}(s,q^2,Q^2)$  appears
when $q^2=0$:
\begin{equation}
\rho^{quark}(s,q^2=0,\, Q^2) = {{Q^2}\over{(s+Q^2)^2}}.
\label{eq:rhoq20}
\end{equation}
This  formula explicitly shows that if
$Q^2$ also  tends to  zero, the spectral density $\rho^{quark}(s,q^2=0,Q^2)$
becomes narrower and higher, approaching  $\delta(s)$ in the
$Q^2 \to 0$ limit (cf. \cite{DZ}).
Thus,  the perturbative triangle diagram
dictates that two real photons can produce only
a single  massless pseudoscalar state:
there are no other states in  the spectrum  of final hadrons.
As $Q^2$ increases, the spectral function broadens,
$i.e.,$ higher states can  also be  produced.
Assuming the local duality, we obtain:
\begin{equation}
f_{\pi} F_{\gamma \gamma^* \pi^0}^{LD}(Q^2)
 = \frac1{\pi } \int_0^{s_0}
\rho^{quark}(s,0,Q^2) \, ds
=  \frac1{\pi (1+Q^2/s_0)} .
\label{eq:FLDgg}
\end{equation}

For large $Q^2$, this gives
\begin{equation}
F_{\gamma \gamma^*  \pi^0 }^{as}(Q^2) =
\frac{4\pi  f_{\pi}}{Q^2} + O(1/Q^4).
\label{eq:fggpias2}
\end{equation}

This result can  also be obtained from the $q^2=0$ version
\begin{equation}
F_{\gamma \gamma^*  \pi^0 }^{pQCD}(Q^2) = \frac{4\pi}{3}
\int_0^1 {{\varphi_{\pi}(x)}\over{xQ^2}} dx  +O(1/Q^4),
\label{eq:gg*pipqcd2}
\end{equation}
of the pQCD formula (\ref{eq:g*g*pipqcd2}),
if we will use there  the asymptotic  form of
the pion distribution amplitude
\begin{equation}
\varphi_{\pi}^{LD}(x) = 6 f_{\pi} x(1-x)
\label{eq:phipiLD}
\end{equation}
produced  by  the local duality prescription.

In other words,   the local duality  formula (\ref{eq:FLDgg})
exactly reproduces the  Brodsky-Lepage interpolation (\ref{eq:blin})
between the $Q^2=0$ value $1/\pi f_{\pi}$  fixed by the ABJ anomaly
and the leading large-$Q^2$ term ${4\pi  f_{\pi}}/{Q^2}$
calculated for the  asymptotic form of the pion distribution amplitude.

\section{Normalization properties of the effective wave function}

\setcounter{equation} 0

\subsection{Momentum representation}

By construction,  $\Psi^{LD}(x,k_{\perp})$  satisfies the standard
constraint
\begin{equation}
\int_0^1 dx \,  \int\frac{ d^2 k_{\perp}}{16\pi^3} \Psi^{LD}(x,k_{\perp}) =
\frac{f_\pi}{2\sqrt{6}}
\label{eq:fpinorm}
\end{equation}
imposed on the two-body Fock component of the pion light-cone wave function
by the correspondence with the $\pi \to \mu \nu$ rate.

Furthermore,  the $x$-integral of  $\Psi^{LD}(x,k_{\perp})$
at zero transverse momentum
\begin{equation}
\int_0^1 dx\,  \Psi^{LD}(x,k_{\perp}=0) = \frac{2\sqrt{6}}{f_\pi}
\label{eq:piggnorm}
\end{equation}
has the right  magnitude to  produce the correct value
$F_{\gamma^* \gamma^* \pi^0}(0,0) =1/\pi f_{\pi}$  imposed by the
$\pi^0 \to \gamma \gamma $ rate.
Note, that  this value is by a factor of 2 larger than the constraint
imposed in
\cite{bhl} on the quark-antiquark component of the LC pion wave function.
The difference  can be traced to  the claim, made in   \cite{bhl},
that the $\bar q q$-component of their
pion wave function gives only a half of the $\pi^0 \to \gamma \gamma$
decay amplitude. The other half, it was  argued,
should be attributed to the $\bar qq \gamma$-component
of the pion wave function.
Within our  approach,
an analogue of the $\bar qq \gamma$-component
appears  only in the first order in $\alpha_s$.
At  the leading order,  there is only one term,
which, I repeat, correctly  reproduces the $Q^2=0$ value of
$F_{\gamma^* \gamma^* \pi^0}(0,Q^2)$.

\subsection{Probability integral}

 The integral
\begin{equation}
P = \int \frac{dx\, d^2 k_{\perp}}{16\pi^3} |\Psi(x,k_{\perp})|^2 .
\label{eq:probab2}
\end{equation}
gives  the probability
to find the pion in the state described by the
wave function $\Psi(x,k_{\perp})$.
As discussed earlier,  substituting  the
local duality wave function $\Psi^{LD}(x,k_{\perp})$ into this relation,
one would get $P = {s_0}/{(4 \pi^2 f_\pi^2)} =1$.
At the lowest order in $\alpha_s$,
the local duality wave function $\Psi^{LD}(x,k_{\perp})$,
in this sense, describes 100\% of the pion content, $i.e,$
 $\Psi^{LD}(x,k_{\perp})$ can  be understood
as an effective wave function  dual to
all Fock components of the pion light-cone wave function.

It is worth noting here that, in  our approach,  $P \leq 1$
for any wave function   $\Psi(x,k_{\perp})$  which \\
$a)$ depends only on
the combination $k_{\perp}^2/x \bar x$:
$\Psi(x,k_{\perp}) = f(k_{\perp}^2/x \bar x)$, \\
$b)$ monotonically decreases with the increase of
$k_{\perp}^2/x \bar x$, \\
$c)$ never becomes negative and \\ $d)$ satisfies
the constraints (\ref{eq:fpinorm}) and (\ref{eq:piggnorm}). \\
The upper limit for $P$ is reached when $\Psi(x,k_{\perp})$
assumes the steplike form (\ref{eq:PsiLD})
dictated by the local duality.
The requirement that the (generalized)
valence content of the pion
should not exceed $100 \%$ is not unreasonable.
Furthermore, I fail to  see
why,  in a particular model, this probability
cannot reach $100\%$.
However, if instead of our constraint (\ref{eq:piggnorm}) one  apllies  that
proposed in \cite{bhl}, the upper limit for $P$,
under the same conditions $a) - d)$,  is 0.5,
$i.e.,$ one should mandatorily require that at least $50\%$
of the pion content must  always be related to non-valence components.

\subsection{Impact parameter representation}

Defining the impact parameter $b_{\perp}$ as  the variable which is
Fourier-conjugate to the transverse momentum $k_{\perp}$:
\begin{equation}
 \Psi(x,k_{\perp}) = \int e^{i k_{\perp} b_{\perp}}
\widetilde \Psi (x,b_{\perp})
d^2 b_{\perp}  \, ,
\end{equation}
we can write down the normalization conditions for the
$b_{\perp}$-space wave function $\widetilde \Psi (x,b_{\perp})$.
Eq. (\ref{eq:fpinorm}), following from the requirement that
$\pi \to \mu \nu$ rate  is specified by $f_{\pi}$,  gives  the magnitude of
$\widetilde \Psi^{LD}(x,b_{\perp}  )$ at the origin:
\begin{equation}
\int_0^1 dx \,  \widetilde \Psi^{LD}(x,b_{\perp} =0 ) =
\frac{2 \pi f_\pi}{\sqrt{6}},
\label{eq:bfpinorm}
\end{equation}
and eq.(\ref{eq:piggnorm}), following from the requirement that
 $\pi^0 \to \gamma \gamma$ rate is given by axial anomaly, specifies
 its integral over
the $whole$ $b_{\perp}$-plane:
\begin{equation}
\int_0^1 dx \, \int  \widetilde \Psi^{LD}(x,b_{\perp} ) \, d^2 b_{\perp}
 = \frac{2\sqrt{6}}{f_\pi}.
\label{eq:b2}
\end{equation}

There is a widespread opinion  that the axial
anomaly is  a  purely short-distance phenomenon
produced by  ultraviolet divergences.
However,  the constraint (\ref{eq:b2})
involving integration over all impact parameters
clearly shows that the axial anomaly is deeply
related to the
long-distance physics as well.
In particular, calculating the spectral density
$\rho(s,q^2,Q^2)$ exhibiting the
anomaly behaviour in the $q^2,Q^2 \to 0$ limit,
we never faced any ultraviolet divergences (cf. \cite{DZ}).

For reference purposes,
we also give the impact-parameter version of the
formula  for the  pion electromagnetic form factor:
\begin{equation}
F_{\pi}^{LD}(Q^2) = \frac1{4\pi} \int_0^1 \,  dx \, \int
e^{ix(  Q_{\perp} b_{\perp})}  |\widetilde \Psi^{LD} (x, b_{\perp}) |^2  \,
d^2 b_{\perp}
\label{eq:bDYW}
\end{equation}
and  the $b_{\perp}$-space form of our  effective
wave function:
\begin{equation}
\widetilde \Psi^{LD} (x, b_{\perp}) = {{\sqrt{6}}\over{\pi f_{\pi} b_{\perp}}}
\sqrt{x \bar x s_0} \,
J_1(b_{\perp} \sqrt{x \bar x  s_0 }),
\label{eq:bWF}
\end{equation}
where $J_1(z)$ is the Bessel function.

\section{Higher-order corrections}

\setcounter{equation} 0

\begin{figure}[t]
\mbox{
   \epsfxsize=12cm
 \epsfysize=3cm
  \epsffile{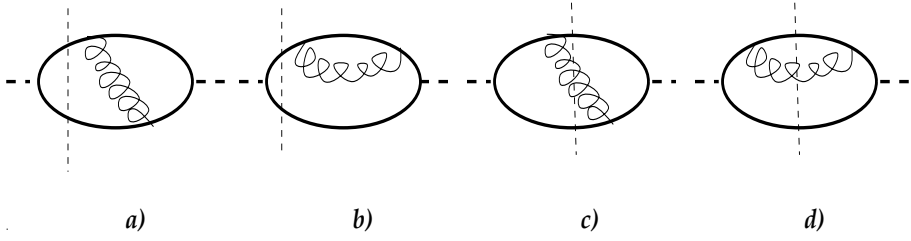}  }
  \vspace{1cm}
{\caption{\label{fig:4}
Some two-loop contributions to  the two-point spectral function
$\rho(s)$ (8.1).
{\it a,b)} Contributions corresponding to $O(\alpha_s)$
corrections to the two-body $\bar q q$
effective wave function.
 {\it c,d)}  Contributions corresponding to presence of hard
three-body $\bar q G q$-components  both in the initial and final
states.}}
\end{figure}

Calculating the spectral densities $\rho^{quark}(s, \ldots)$
to higher orders
in $\alpha_s$, we can study effects due to gluon radiation.
Depending on the position of Cutkosky cuts,
one can interpret, $e.g.,$  the next-to-leading order contributions
either as corrections to the two-body $\bar q q$
effective wave function (Fig.\ref{fig:4}$a,b$)
or as three-body  $\bar q Gq$ Fock components (Fig.\ref{fig:4}$c,d$) .

In practice,  even the lowest $O(\alpha_s)$ correction
requires a two-loop calculation, which is rather involved,
especially for three-point functions.
For the two-point function, the correction is known \cite{svz}:
\begin{equation}
\rho^{quark}_{NLO}(s) \equiv \rho^{quark}_{0}(s) + \rho^{quark}_{1}(s) =
 \frac1{4\pi}\left (1+\frac{\alpha_s}{\pi} \right ).
\label{eq:ronlo}
\end{equation}
According to the Ward identity (\ref{eq:WID}),  this result can be used to
get the $Q^2 =0$ value of the  $O(\alpha_s)$ contribution  to  the
spectral density $\rho^{quark}(s_1,s_2,Q^2)$
related to the pion  electromagnetic form factor.
As a result,  the $O(\alpha_s)$-correction
to the pion form factor
for $Q^2 =0$  is given by
\begin{equation}
\delta F_{\pi}^{(\alpha_s)} (Q^2 = 0) = \frac{\alpha_s(s_0)}{\pi}.
\end{equation}
The duality interval $s_0$, in this case, is a natural (and the only possible)
scale for the running coupling constant.

\begin{figure}[t]
\mbox{
   \epsfxsize=12cm
 \epsfysize=4cm
  \epsffile{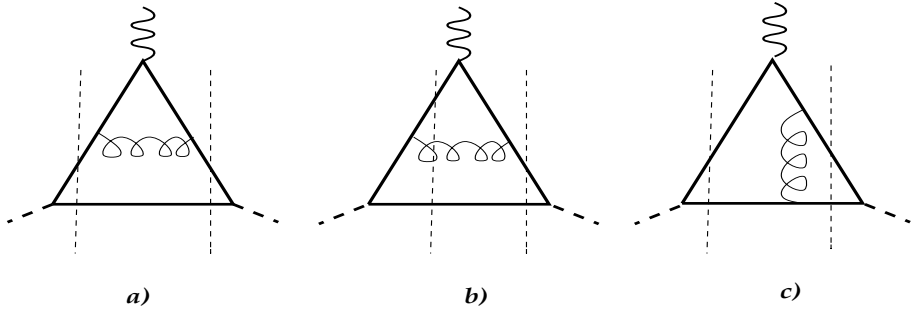}  }
  \vspace{1cm}
{\caption{\label{fig:5}
Some two-loop contributions for the spectral function
$\rho(s_1,s_2,Q^2)$.  {\it a)} Correction to the electromagnetic vertex.
{\it b)} Contribution corresponding to the
three-body $\bar q G q$-component  in the initial state and
two-body $\bar q  q$-component in the final
state. {\it c)} Term producing the $O(1/Q^2)$ contribution. }}
\end{figure}

Another important piece of information  can be obtained
for large $Q^2$. In this limit, in contrast to the one-loop term
$\rho^{quark}_0(s_1,s_2,Q^2)$,  which  decreases like $1/Q^4$,
the two-loop contribution  $\rho^{quark}_1(s_1,s_2,Q^2)$
contains a term (Fig.\ref{fig:5}$c$) which behaves like  $1/Q^2$:
\begin{equation}
\rho^{quark}_1(s_1,s_2,Q^2) = \frac{8 \pi \alpha_s}{Q^2}
\rho_0^{quark}(s_1) \rho_0^{quark}(s_2) + O(1/Q^4),
\end{equation}
where $\rho^{quark}_0(s_1)$ and   $\rho^{quark}_0(s_2)$ are
the lowest-order two-point  function spectral densities (see
eq.(\ref{eq:ronlo})).
This behaviour agrees with
the pQCD factorization theorem \cite{tmf78,bl80} and quark counting rules.
Substituting  the asymptotic expression for
$\rho^{quark}_1(s_1,s_2,Q^2)$ into the local duality relation
(\ref{eq:FpiLD}),
we get the large-$Q^2$ behaviour of $\delta F_{\pi}^{(\alpha_s)} (Q^2)$:
\begin{equation}
\delta F_{\pi}^{(\alpha_s)} (Q^2) =   \frac{\alpha_s(s_0)}{\pi}
\left ( \frac {2s_0}{Q^2} \right ).
\end{equation}
This result  corresponds to the pQCD  formula for the
one-gluon-exchange contribution to the pion form factor
\cite{ar77,tmf78,blpi79}
\begin{equation}
F_\pi^{pQCD}(Q^2)=\int_0^1 dx\int_0^1 dy\
\varphi_\pi(x) \ \varphi_\pi(y)  \frac{8\pi \alpha_s }{9xyQ^2} \, ,
   \label{eq:qcdff}
\end{equation}
if one uses  the ``asymptotic'' distribution amplitude
$ \varphi_\pi^{as} (x) = 6 f_{\pi} x (1-x)$
dictated by the local duality.
Now, by analogy with the Brodsky-Lepage interpolation,
we can construct  a model for $\delta F_{\pi}^{(\alpha_s)} (Q^2)$
based on the simplest  interpolation between its
$Q^2 = 0 $ value and large-$Q^2$  asymptotics:
\begin{equation}
 \delta F_{\pi}^{(\alpha_s)} (Q^2) = \left ( \frac{\alpha_s}{\pi }\right )
\frac1{1+Q^2/2s_0}.
\end{equation}
Combining the $O(1)$ and  $O(\alpha_s)$   terms, we get
the next-to-leading order LD-model  for the pion form factor
\begin{equation}
F_{\pi}^{LD} (Q^2) = {{F_{\pi}^{LD(0)}(Q^2) +
\delta F_{\pi}^{(\alpha_s)}  (Q^2)}\over{1+ \frac{\alpha_s}{\pi }}}.
\label{eq:FpiNLO}
\end{equation}
where $F_{\pi}^{LD(0)}(Q^2)$ is the lowest-order result given by
eq.(\ref{eq:sqrt}).
For $\alpha_s/\pi$ one can take a constant value
$\alpha_s/\pi = \alpha_s(s_0)/\pi \approx 0.1$
though, for truly asymptotic $Q^2$,
the scale of $\alpha_s$  should have a $Q^2$-dependent component.
The  curve based on eq.(\ref{eq:FpiNLO})  is in good agreement with existing
data.

\section{Conclusions}

\setcounter{equation} 0

Our main goal here was to demonstrate that the results of
the approach  based on local quark-hadron duality
and QCD sum rules
can be reformulated in terms of the effective wave function
$\Psi^{LD}(x,k_{\perp})$ describing both longitudinal
and transverse momentum distribution of quarks inside the pion.
This approach has  the following features: \\
$1)$ It is directly related to the QCD Lagrangian,
and all the calculations are based on Feynman diagrams
of standard 4-dimensional perturbation theory. \\
$2)$ As a result, the approach is fully compatible with high-$Q^2$
pQCD calculations and other QCD constraints, $e.g.,$ those
imposed by the axial anomaly. \\
$3)$  Radiative (higher-order in $\alpha_s$) corrections
can be added in a regular way, through a well-defined procedure.\\
$4)$ There is no need for a special procedure
separating soft $vs.$ hard terms.
In a sense, they  are separated automatically
by the duality interval parameter $s_0$.
The ``hard'' terms  have a natural subasymptotic modification
in the low-$Q^2$ region. \\
$5)$ The bulk (soft) part of the higher-twist
effects is  described by an effective
2-body wave function $\Psi^{LD}(x,k_{\perp})$  rather than
by increasingly complex wave functions of  higher Fock
components. \\
$6)$ In this approach, the  contributions which can be interpreted
as effective wave functions for the  higher Fock components
are small because they are suppressed
by  powers of  $\alpha_s(s_0)/\pi$. Hence,  the
effective valence component dominates and, in this respect,
this approach  resembles   the constituent quark model.
However, there is no need
to introduce  constituent quark masses.
The scale responsible for the IR cut-offs
is  set by the duality interval $s_0$.  \\
$7)$ The effective  wave functions are introduced
in this approach  without any appeal to the existence of
bound state equations.

Local quark-hadron duality was also applied to
nucleon form factors \cite{nr83} and to $\gamma p \to \Delta$
transition form factors \cite{belrad}.
The results of these studies can be used
 to develop a similar
formalism for the baryons.
Another possible development is to substitute  the
steplike effective wave functions
 $\Psi^{LD}(x,k_{\perp})$ by smooth functions, but
without violating the constraints (\ref{eq:fpinorm}),
(\ref{eq:piggnorm})  and $P^{(0)} =1$.

\section{Acknowledgement}

I am most grateful to V.M.Braun, M.Veltman, F.J. Yndurain and
V.I.Zakharov
for stimulating criticism,
 to S.J.Brodsky
for a  correspondence  about the BHL prescription
and to V.M.Belyaev, W.Broniowski,
C.E. Carlson, F.Gross, L.L.Frankfurt, N.Isgur,
L.Mankiewicz, I.V. Musatov,
M.A. Strikman  and  A.R.Zhitnitsky
for useful discussions.
I would like to thank A.Bialas, W.Czy\.z, W.Broniowski, M.Nowak,
 M.Praszalowicz  and J.Szwed for   kind hospitality
in  Cracow and at XXXV Cracow School in Zakopane.
This work was supported by the
US Department of Energy  under contract DE-AC05-84ER40150 and by
Polish-U.S. II Joint Maria Sklodowska-Curie
Fund, project number PAA/NSF-94-158.

\end{document}